# Development of the Reiss theory for binary homogeneous nucleation of aerosols

M.S. Veshchunov[*)]

Nuclear Safety Institute (IBRAE), Russian Academy of Sciences,

52, B. Tulskaya, Moscow 115191, Russian Federation

**Abstract**

The Reiss theory for binary homogeneous nucleation in binary gas mixtures is critically analysed and further developed. Based on the analysis of phase space trajectories in the supercritical zone of the phase transition, carried out within the framework of the theory of two-dimensional dynamical systems and supplemented by the flux matching condition at the boundary of the critical zone, it is shown how the theory should be modified. The proposed modification is equivalent to the earlier modifications by Langer and Stauffer, based on additional trial assumptions (ansatz) for solving the steady state equation for the non-equilibrium size distribution function, but reveals and substantiates the approximation underlying their approach. The extension of the Reiss theory to binary vapours in inert carrier (atmospheric) gases is justified.

[*)] Corresponding author. E-mail: msvesh@gmail.com

## 1. Introduction

Condensation of water vapour, leading to the formation of clouds in the atmosphere, occurs due to supersaturation of water. In the atmosphere, a large supersaturation is prevented by the presence of abundant condensation nuclei. The main sources of such nuclei are insoluble particles from smoke and dust, soluble particles such as sodium chloride from the oceans or ammonium sulfate from a chemical reaction in the atmosphere, and pre-existing suspended liquid particles (liquid aerosols). Liquid aerosols are of particular importance as they are known to be one of the major components of photochemical smog. These liquid aerosols can be generated in the atmosphere by the binary homogeneous nucleation of water vapour and a polluting reactant such as sulfuric acid or nitric acid [1, 2].

The problem of nucleation rate of these liquid aerosols is an example of homogeneous nucleation in binary systems where the nuclei can be considered to consist of spherical droplets of a new phase. However, the classical nucleation theory [3–5] was developed mainly in relation to one-component (unary) systems. In accordance with this theory, the point of unstable equilibrium between a cluster of a new phase and a surrounding parent phase corresponds to a maximum of the free energy $\Delta G_0(x)$ (where $x$ is the cluster size) representing the reversible work of formation of the cluster.

This theory was generalized to the kinetics of nucleation in binary mixtures by Reiss [6]. In the Reiss theory, the parent phase is thought of as consisting of molecules of two components $X$ and $Y$, together with clusters of the new phase. A particular molecular cluster is characterized by the numbers of single molecules (or monomers) $x$ and $y$ of species $X$ and $Y$, respectively, that it contains. Reiss showed that the point of unstable equilibrium corresponds in this case to a saddle point on the free energy surface $\Delta G_0(x, y)$. He characterized the rate of the transition by a two-dimensional flux vector (in the phase space of cluster sizes) $\boldsymbol{J}(x, y)$ oriented in the direction of the steepest descent of the free energy surface; for this reason, this orientation was determined in a purely thermodynamic approach.

In the present work, the Reiss theory will be analysed and developed by more accurate consideration of the steady state flow of clusters through the critical zone of the phase space based on the analysis of the phase space trajectories in the supercritical zone, carried out within the framework of the theory of two-dimensional dynamical systems and supplemented by the flux matching condition at the boundary of the critical zone. This requires a modification of the theory associated with a different direction of the flow vector determined by the basic axis of the linearized system of equations for the cluster growth rate $(\dot{x}, \dot{y})$ in the supercritical zone.

This modification turns out to coincide with the modification proposed by Langer [7] and Stauffer [8] using a more formal approach based on additional trial assumptions ('ansatz') to solve the stationary equation for the non-equilibrium size distribution function in the critical zone; however, our analysis will reveal and justify the approximation underlying their approach.

The Reiss theory, developed only for pure binary gas mixtures in the absence of a third inert gas, was applied (without justification) in subsequent works to the binary nucleation of liquid aerosols in the atmosphere, using the same expression for the equilibrium size distribution function. Such approach, which was widely criticised in the literature, will receive further justification. This will avoid using alternative expressions for the pre-exponential factor in the nucleation rate from the literature.

## 2. Analysis of the Reiss theory

Generalizing the classical nucleation theory for unary systems [3–5], Reiss related the nucleation rate in binary $(x, y)$ systems to the steady state flux of clusters in the phase space, $\boldsymbol{J}(x, y)$, which in the continuous approximation for the size distribution function $f(x, y)$ may be written as ([6])

$$J_x = -\beta_x f_0 \frac{\partial}{\partial x}\left(\frac{f}{f_0}\right) = -\beta_x \frac{\partial f}{\partial x} + \left(\beta_x \frac{1}{f_0}\frac{\partial f_0}{\partial x}\right)f, \qquad (1)$$

$$J_y = -\beta_y f_0 \frac{\partial}{\partial y}\left(\frac{f}{f_0}\right) = -\beta_y \frac{\partial f}{\partial y} + \left(\beta_y \frac{1}{f_0}\frac{\partial f_0}{\partial y}\right)f, \qquad (2)$$

where $\beta_x(x,y)$ and $\beta_y(x,y)$ are the arrival rates of monomers $X$ and $Y$, respectively, into cluster $(x,y)$; the equilibrium size distribution function has the form

$$f_0(x,y) = F \exp[-\Delta G_0(x,y)/kT], \qquad (3)$$

$\Delta G_0(x,y)$ is the Gibbs free energy of cluster formation, and the pre-exponential factor $F$ will be discussed below in Section 5. He noted that the flow of clusters is not confined to a path through the saddle point alone, but may occur across a narrow ridge passing through the saddle, and this flow was integrated in order to arrive at the rate of nucleation.

Reiss' consideration was based on two assumptions: 1) the steady state flow of clusters through the critical zone is so pronounced in one direction that, in comparison with it, any lateral flow may be neglected; and 2) the axis of the pass $x'$ runs in the direction of the steepest descent of the free energy surface $\Delta G_0(x,y)$, which for this reason is determined in the thermodynamic approach (i.e. solely from the properties of the free energy).

Denoting the component of $\mathbf{J}$ parallel to the axis of the flow $x'$ by $J_{x'}$, and the component in the perpendicular direction $y'$ by $J_{y'}$, which obey the steady state condition,

$$\mathrm{div}\mathbf{J} = \frac{\partial J_{x'}}{\partial x'} + \frac{\partial J_{y'}}{\partial y'} = 0, \qquad (4)$$

the first assumption amounts to the assertion $J_{y'} = 0$, which when substituted into Eq. (4) yields

$$\frac{\partial J_{x'}}{\partial x'} = 0, \qquad (5)$$

i.e. the flux $J_{x'} = J(y')$ is a function of $y'$ only [6].

As for the second assumption of the Reiss theory, it may be incompatible with the first, as shown below.

## 3. Two-dimensional dynamical system analysis

In accordance with the Reiss theory, the point of unstable equilibrium corresponds to a saddle point on the free energy surface $\Delta G_0(x,y)$, i.e. a point at which the slopes (or derivatives) in orthogonal directions are all zero, while the curvature of the surface is negative in one direction and positive in the perpendicular direction. This conclusion was confirmed by Maydet and Russel [9], who noted that the critical nucleus of size $(x^*, y^*)$ can be determined from the steady state condition for an individual cluster, represented as $\dot{x}|_{x^*,y^*} = \dot{y}|_{x^*,y^*} = 0$. Accordingly, in the phase space of cluster size $(x,y)$, where trajectories $y(x)$ are solutions of the cluster growth rate equations, the critical point is a saddle determined by the intersection of two nodal lines, $\dot{x} = 0$ and $\dot{y} = 0$, as shown in Fig. 1.

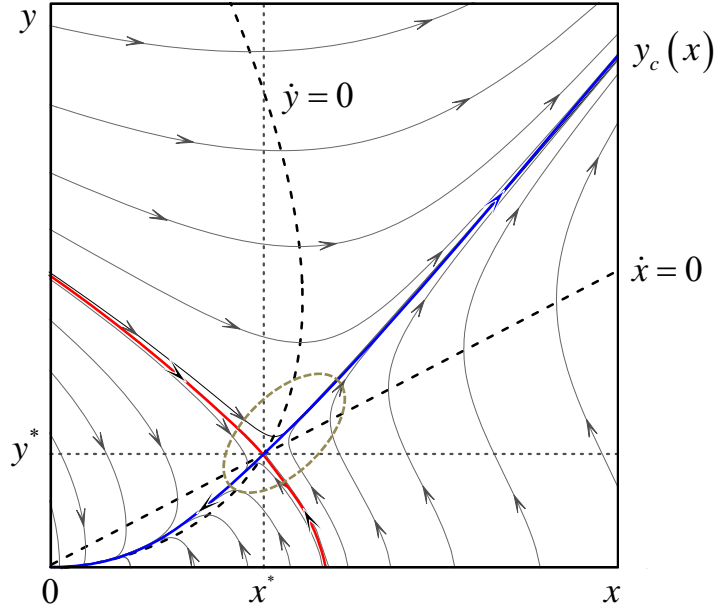

Fig. 1. Illustration (after [9]) of the critical (saddle) point $(x^*, y^*)$, nodal lines (dashed lines) and cluster trajectories (solid lines) with arrows indicating direction of the phase velocity vector $\mathbf{v} = (v_x, v_y)$ defined in Eqs (9) and (10); the ellipsoid surrounding the saddle point schematically represents the critical zone.

For a one-component system with the Gibbs free energy of cluster formation $\Delta G_0(x)$, the growth kinetics of large clusters, $x \gg 1$ (to which the classical nucleation theory is applicable), can be described as

$$\dot{x} = v_x(x) = \beta_x - \alpha_x = \beta_x\left(1 - \frac{\alpha_x}{\beta_x}\right) = \beta_x\left[1 - \exp\left(\frac{1}{kT}\frac{d\Delta G_0(x)}{dx}\right)\right], \tag{6}$$

where $\beta_x = \frac{PS_x}{(2\pi mkT)^{1/2}}$ is the arrival rate of vapour molecules (with the partial pressure $P$, molecular mass $m$ and volume $\Omega$) to a sperical cluster of radius $R_x = (3\Omega x/4\pi)^{\frac{1}{3}}$ and surface area $S_x = 4\pi R_x^2 = (36\pi\Omega^2)^{\frac{1}{3}}x^{\frac{2}{3}}$; $\alpha_x = \frac{P_x^{(0)}S_x}{(2\pi mkT)^{1/2}}$ is the evaporation rate from the cluster, $P_x^{(0)}$ is the equilibrium partial pressure over the cluster.

Indeed, representing the formation free energy in the form, $\Delta G_0(x) = -xkT\ln(P/P_0) + \sigma S_x$, where $\sigma$ is surface tension, $P_0$ is the saturation pressure, and calculating $P_x^{(0)}$ using the Gibbs-Kelvin equilibrium condition, $d\Delta G_0(x)/dx = 0$ (see, e.g. [10]), one obtains $kT\ln\left(P_x^{(0)}/P_0\right) = \sigma\, dS_x/dx$, which leads to the relation $\frac{\alpha_x}{\beta_x} = \frac{P_x^{(0)}}{P} = \frac{P_0}{P}\exp\left(\frac{1}{kT}\frac{\sigma dS_x}{dx}\right) = \exp\left(\frac{1}{kT}\frac{d\Delta G_0(x)}{dx}\right)$, used in Eq. (6).

At the critical point, $d\Delta G_0(x)/dx = 0$ (and $P_{x^*}^{(0)} = P$), and hence Eq. (6) can be simplified in a small vicinity of the critical point, where $d\Delta G_0(x)/dx \ll 1$, as

$$\dot{x} \approx -\frac{\beta_x}{kT}\frac{\partial \Delta G_0(x)}{\partial x}, \tag{7}$$

and further linearized as

$$\dot{x} = -\frac{\beta_x^*}{kT}\left[\frac{\partial^2 \Delta G_0(x^*)}{\partial x^2}(x - x^*) + \frac{1}{3}\frac{\partial^3 \Delta G_0(x^*)}{\partial x^3}(x - x^*)^2 + \cdots\right] \approx -\frac{\beta_x^*}{kT}\frac{\partial^2 \Delta G_0(x^*)}{\partial x^2}(x - x^*), \tag{8}$$

where $\beta_x^* = \beta_x(x^*)$. This expansion takes into account that $\Delta G_0^{(3)}(x^*) \equiv \frac{\partial^3 \Delta G_0(x^*)}{\partial x^3} = \frac{\partial^3}{\partial x^3} \sigma(36\pi\Omega^2)^{\frac{1}{3}} x^{2/3}\Big|_{x=x^*} = -\frac{4}{3} x^{*-1} \Delta G_0^{(2)}(x^*)$, or, more generally, for higher derivatives of order $k \geq 2$, $\Delta G_0^{(k+1)}(x^*) = x^{*-1} O\left(\Delta G_0^{(k)}(x^*)\right)$, and hence the higher order terms in Eq. (8) can be neglected if $|x - x| \ll x^*$.

The generalization of this derivation to binary systems with the Gibbs free energy $\Delta G_0(x, y)$ is straightforward and leads to

$$\dot{x} = v_x(x, y) = \beta_x - \alpha_x = \beta_x \left[1 - \exp\left(\frac{1}{kT} \frac{\partial \Delta G_0(x,y)}{\partial x}\right)\right] \approx -\beta_x \frac{1}{kT} \frac{\partial \Delta G_0(x,y)}{\partial x}, \tag{9}$$

$$\dot{y} = v_y(x, y) = \beta_y - \alpha_y = \beta_y \left[1 - \exp\left(\frac{1}{kT} \frac{\partial \Delta G_0(x,y)}{\partial y}\right)\right] \approx -\beta_y \frac{1}{kT} \frac{\partial \Delta G_0(x,y)}{\partial y}. \tag{10}$$

Similar equations were obtained in [8, 9] from discrete detailed balancing equations, assuming that for large clusters they can be regarded as continuous.

Therefore, the steady state condition, $\dot{x} = \dot{y} = 0$, corresponds to the equation $\frac{\partial \Delta G_0(x,y)}{\partial x} = \frac{\partial \Delta G_0(x,y)}{\partial y} = 0$, which is valid in the critical point $(x^*, y^*)$, in agreement with Reiss' thermodynamic analysis. In a small vicinity of the critical point $|x_i - x_i^*| \ll x_i^*$, Eqs (9) and (10) can be linearized similarly to Eq. (8) as

$$\dot{x} \approx -\frac{\beta_x^*}{kT} \left[\frac{\partial^2 \Delta G_0(x^*,y^*)}{\partial x^2}(x - x^*) + \frac{\partial^2 \Delta G_0(x^*,y^*)}{\partial x \partial y}\Big|_{x^*,y^*} (y - y^*)\right], \tag{11}$$

$$\dot{y} \approx -\frac{\beta_y^*}{kT} \left[\frac{\partial^2 \Delta G_0(x^*,y^*)}{\partial x \partial y}(x - x^*) + \frac{\partial^2 \Delta G_0(x^*,y^*)}{\partial y^2}(y - y^*)\right], \tag{12}$$

where $\beta_{x,y}^* = \beta_{x,y}(x^*, y^*)$.

In accordance with the theory of two-dimensional dynamical systems, there are two basic axes which intersect at the saddle point, where they are directed along two eigenvectors of the linearized system of Eqs (11) and (12) (cf. Chapter 3.1 of [11]). In this type of phase portrait, the two trajectories given by the eigenvectors of the negative eigenvalue of the linearized system of Eqs (11) and (12) move toward and eventually converge at the critical point (stable line). The two trajectories given by the eigenvectors of the positive eigenvalue move in exactly the opposite way: start at the critical point and move away (unstable line). Every other trajectory $y(x)$ moves toward but never converges to the critical point, before changing direction and then moves away. The tangent of each trajectory determines the direction (indicated by arrows in Fig. 1) of phase velocity vector $\boldsymbol{v} = (v_x, v_y)$ with the modulus $v = dy(x)/dx = \dot{y}/\dot{x}$ at each point $(x, y)$.

The two trajectories given by the eigenvectors of the negative eigenvalue (the first basic axis, red line in Fig. 1) separate the phase space into two parts, in which single clusters grow or dissolve respectively. The two trajectories given by the eigenvectors of the positive eigenvalue (the second basic axis, blue line in Fig. 1) describe the evolution of a cluster of the critical size in the sub- and supercritical zone, respectively. Any other trajectory is approaching this basic axis either in the subcritical or supercritical zone.

Accordingly, since the critical point $(x^*, y^*)$ in Eq. (12) is a saddle, the determinant of the matrix $\mathbf{D} = (D_{ij})$ with elements $D_{ij} = \frac{1}{2} \frac{\partial^2 \Delta G_0(x^*,y^*)}{\partial x_i \partial x_j}$ is negative (cf. Chapter 4.1 of [11]), $\det \mathbf{D} < 0$, while the second basic axis (blue line) is directed along the eigenvector with the positive eigenvalue of the matrix $-\boldsymbol{\beta}\mathbf{D}$ (or the negative eigenvalue of the matrix $\boldsymbol{\beta}\mathbf{D}$), where $\boldsymbol{\beta} = \begin{pmatrix} \beta_x & 0 \\ 0 & \beta_y \end{pmatrix}$.

## 4. Modification of the Reiss theory

As noted in [5], in the macroscopic theory, each nucleus changes with time in a quite definite way, depending on its size $(x, y)$ and external conditions; $N$ nuclei which were identical at the initial moment will remain identical even after a certain time interval, and will be shifted as a whole to another place of the phase space that corresponds to the change in the size of the nucleus, in accordance with macroscopic equations. Therefore, in the supercritical zone, where the growth of a cluster takes place in accordance with macroscopic Eqs (9) and (10), the flux of clusters takes the form $\bar{J} = vf$, where $f(x, y)$ is the size distribution function and $v(x, y)$ is the phase velocity vector (defined above).

This conclusion can be confirmed directly by substituting Eq. (3) into Eqs (1) and (2) and using Eqs (9) and (10) for $v_x, v_y$, which leads to

$$J_i = -D_i \frac{\partial f}{\partial x_i} + v_i f, \quad i = x, y, \tag{13}$$

where the first term in the r.h.s. describes the 'diffusion' part of the steady state flux (with $D_i = \beta_i$) in the phase space of cluster size $(x, y)$, which is related to fluctuations (significant in the critical zone and gradually decreasing towards its periphery), whereas the second term describes the 'advection' part, $\left(\beta_i \frac{1}{f_0} \frac{\partial f_0}{\partial x_i}\right) f = -\left(\beta_i \frac{1}{kT} \frac{\partial G_0}{\partial x_i}\right) f = v_i f$ (significant both in critical and supercritical zones). Correspondingly, outside the critical zone, where fluctuations can be neglected, only the advection part exists, $\bar{J} = vf$, which smoothly transforms into the complete Eq. (13) at the periphery of the critical zone.

Since the passage ridge is narrow (due to the exponential decrease in the cluster flux with the growth of the free energy barrier in the vicinity of the saddle point), off-critical clusters that have penetrated the critical zone, move in the supercritical zone along trajectories that are approaching to and line up in the direction of the second basic axis (blue line), which describes the trajectory of the critical cluster (see Fig. 1). Hence it follows that that the orientation of the flux $\bar{J}$ in the supercritical zone predominates in the direction of the second basic axis. Since the fluxes in the critical and supercritical zones must match with each other at the boundary of the critical zone, the flow in the critical zone also predominates in the same direction. This asserts that in the critical zone, where the velocity component perpendicular to the second basic axis is not small, the advection part of the flux component in this direction is well balanced by its diffusion part, which is also not small there. This conclusion is consistent with Reiss' first assumption, but contradicts the second.

A similar conclusion was derived by Langer [7] for multicomponent systems (characterized by the multidimensional phase space) more formally, using a trial assumption (ansatz) that the function $f/f_0$ is a function of one variable $u = U_1 x + U_2 y$ (here, for simplicity, being reduced to a binary formulation). Based on this assumption, he showed that the flux vector $J$ in the critical zone is constant and parallel to the direction of the unstable mode at the saddle point.

However, it can be seen from the above consideration that Langer's derivation holds only approximately, i.e. the flow dominates in the direction of the second basic axis, but is not exactly parallel to it. Indeed, at the periphery of the critical zone (near its boundary with the supercritical zone), where the advection part of the flux prevails over the diffusion (fluctuation) part, the flux component perpendicular to the second basic axis is relatively small but does not vanish, see Fig. 1. This means that Langer's trial assumption is not a strict relationship (as is commonly believed) but only an approximation, which can be justified from the flux matching condition proposed in the present approach.

Later, similar results were obtained by Stauffer [8] (but without reference to [7]), assuming (following Reiss) that the direction of the flux is constant in the critical zone, $J_{x'} = J(y'), J_{y'} = 0$, and using another trial assumption for the flux, $J_{x'} \propto \exp(-W y'^2)$, with a width factor $W$ determined in further calculations. However, both of these assumptions were derived in [7] directly from Langer's ansatz and thus naturally led Stauffer to the same conclusion regarding the flux direction. This demonstrates the equivalence of both approaches [7] and [8], which was additionally confirmed in [12].

Accordingly, some modification of the Reiss theory is required, associated with a different direction of the axis of the pass, determined from the kinetic consideration (i.e. from rate equations, Eqs (11) and (12)), rather than from the thermodynamic consideration of the steepest descent of the free energy surface (used by Reiss). This modification is related to the new definition of the angle $\theta$ between the original axis $x$ and the axis of the pass $x'$, which coincides with the second basic axis (blue line) of the phase portrait, Fig. 1. Therefore, the angle $\theta$ is given by the equation

$$\tan\theta = \frac{dy_c(x^*)}{dx}, \tag{14}$$

where $y_c(x)$ is the solution of the macroscopic equation

$$\frac{dy}{dx} = \frac{v_y}{v_x} = \beta_y \frac{\partial \Delta G_0(x,y)}{\partial y} \Big/ \left(\beta_x \frac{\partial \Delta G_0(x,y)}{\partial x}\right), \tag{15}$$

which is derived from the system of Eqs (9) and (10) with the boundary condition $y_c(x^*) = y^*$, and describes the critical trajectories through the saddle point.

Therefore, Reiss' expression for the flux

$$J_{x'} = J(y') = \frac{\beta_x^* \beta_y^* (1+\tan^2\theta)}{\beta_y^* + \beta_x^* \tan^2\theta} \left(\int_0^\infty \frac{dx'}{f_0(x',y')}\right)^{-1}, \tag{16}$$

should be modified by using $\theta$ defined in Eq. (14), where $f_0(x',y') = F \exp[-\Delta G_0(x',y')/kT]$. In the quadratic approximation (near the critical point),

$$\Delta G_0(x',y') \approx \Delta G_0^* + P'(x'-x^*)^2 + Q'(y'-y^*)^2 + 2R'(x'-x^*)(y'-y^*), \tag{17}$$

where $\Delta G_0^* = \Delta G_0(x^*, y^*)$, and

$$P' = \frac{1}{2}\frac{\partial^2 \Delta G_0(x^*,y^*)}{\partial x'^2} = P\cos^2\theta + Q\sin^2\theta + 2R\sin\theta\cos\theta, \tag{18a}$$

$$Q' = \frac{1}{2}\frac{\partial^2 \Delta G_0(x^*,y^*)}{\partial y'^2} = P\sin^2\theta + Q\cos^2\theta - 2R\sin\theta\cos\theta, \tag{18b}$$

$$R' = \frac{\partial^2 \Delta G_0(x^*,y^*)}{\partial x' \partial y'} = (Q-P)\sin\theta\cos\theta + R\cos 2\theta, \tag{18c}$$

where $P \equiv D_{11} = \frac{1}{2}\frac{\partial^2 \Delta G_0(x^*,y^*)}{\partial x^2}$, $Q \equiv D_{22} = \frac{1}{2}\frac{\partial^2 \Delta G_0(x^*,y^*)}{\partial y^2}$, $R \equiv D_{12} = \frac{1}{2}\frac{\partial^2 \Delta G_0(x^*,y^*)}{\partial x \partial y}$, are elements of the matrix $\mathbf{D} = \begin{pmatrix} P & R \\ R & Q \end{pmatrix}$ (included in Eqs (11) and (12)) with $\det\mathbf{D} = QP - R^2 = Q'P' - R'^2$, which is invariant with respect to the rotation of the orthogonal coordinate axes due to the invariance of the free energy $\Delta G_0(x, y)$. In the orthogonal system of coordinates associated with the directions of the steepest descent and the steepest ascent at the saddle point, respectively, used by Reiss and denoted hereafter $(x'', y'')$, it can also be represented in the form $\det\mathbf{D} = Q''P'' = -Q''|P''| < 0$.

Under the condition $P' < 0$, there is a maximum at $x = x^*$ of the free energy in the newly defined direction of the $x'$-axis. If $x^*(|P'|/kT)^{\frac{1}{2}} \gg 1$ (when the critical range of nucleus sizes $x$ near the critical size $x^*$, with width $\delta x \sim (kT/|P'|)^{\frac{1}{2}}$, is small compared to $x^*$), this maximum is sufficiently steep to use Eq. (17) in the integrand of Eq. (16) and to extend the integration over $x - x^*$ from $-\infty$ to $\infty$. This leads to the expression

$$J(y' - y^*) = \frac{\beta_x^* \beta_y^* (1+\tan^2 \theta)}{\beta_y^* + \beta_x^* \tan^2 \theta} F \left(\frac{|P'|}{\pi kT}\right)^{1/2} \exp\left(-\frac{\Delta G_0^*}{kT}\right) \exp\left[-\frac{\left(\frac{R'^2}{|P'|} + Q'\right)(y' - y^*)^2}{kT}\right], \tag{19}$$

and thus the total flux (equal to the nucleation rate) is evaluated as

$$J^* \approx \int_{-\infty}^{\infty} J(y') dy' = \frac{\beta_x^* \beta_y^* (1+\tan^2 \theta)}{\beta_y^* + \beta_x^* \tan^2 \theta} F \left(\frac{|P'|}{\pi kT}\right)^{1/2} \left(\frac{\pi kT}{Q' + R'^2/|P'|}\right)^{1/2} \exp\left(-\frac{\Delta G_0^*}{kT}\right) =$$

$$\frac{\beta_x^* \beta_y^* (1+\tan^2 \theta)}{\beta_y^* + \beta_x^* \tan^2 \theta} F |P'| \left(\frac{1}{R^2 - QP}\right)^{1/2} \exp\left(-\frac{\Delta G_0^*}{kT}\right), \tag{20}$$

in agreement with [8], as well as with a more general form for multicomponent systems obtained by Langer [7]

$$J^* = \frac{|k|}{|\det \mathbf{D}|^{1/2}} F \exp\left(-\frac{\Delta G_0^*}{kT}\right), \tag{21}$$

where $k$ is the negative eigenvalue of the matrix $\boldsymbol{\beta}\mathbf{D}$, and $\boldsymbol{\beta}$ is the arrival rate matrix defined for binary systems in Section 3 (see details in [12]).

However, this equation is valid only when the curvature $Q''$ across the saddle point in the direction of the steepest ascent $y''$ is so small that $Q' + R'^2/|P'| = Q''|P''|/|P'| \ll \pi kT$; in this case, the concentration of off-critical nuclei will be appreciable, since the nucleation barrier will be only slightly larger than $\Delta G_0^*$, and many can flow over the barrier. Indeed, the integration in Eq. (20) can be used with confidence instead of the (original) summation, $\sum_{i=-\infty}^{\infty} J(i) = J(0) + \sum_{i=-\infty}^{-1} J(i) + \sum_{i=1}^{\infty} J(i) \approx \int_{-\infty}^{\infty} J(y') dy'$, if $J(0) \ll \sum_{i=-\infty}^{-1} J(i) + \sum_{i=1}^{\infty} J(i)$, which is equivalent in this case to $J(0) \ll \int_{-\infty}^{\infty} J(y') dy'$, or $\left(\frac{\pi kT}{Q' + R'^2/|P'|}\right)^{1/2} \ll 1$, as seen by comparing Eq. (19) (giving $J(0)$) with Eq. (20).

In the opposite case of large curvature $Q''$ (discussed by Reiss [6]), leading to $Q' + R'^2/|P'| \gg \pi kT$, the saddle point passage is narrow and has very steep sides, and only those clusters with the composition of the saddle point nucleate in any noticeable quantity. To take this effect into account, the summation over the cluster sizes in this direction should be restored (instead of the integration in Eq. (20)), which, due to the strong suppression of off-critical nuclei, takes the form

$$J^* = \sum_{i=-\infty}^{\infty} J(i) = \sum_{i=-\infty}^{-1} J(i) + \sum_{i=1}^{\infty} J(i) + J(0) \approx J(0). \tag{22}$$

In this case, the total flow contribution of off-critical nuclei with higher formation barriers, which move along trajectories adjacent to the 'classical' trajectory (blue line), can be considered as a correction, which is neglected in the so called 'quasi-classical' approximation (taking into account only the extreme classical trajectory). In this approximation, the nucleation problem becomes quasi-one-dimensional, by formally substituting $y_c(x)$ into the expression for $\Delta G_0(x, y)$ and further analysing a one-dimensional system with $\Delta G_0(x) = \Delta G_0(x, y_c(x))$ in the framework of the classical nucleation theory. In complex cases, the quasi-classical approximation makes it possible to carry out a simplified (conservative) analysis of binary systems using the one-dimensional classical theory of nucleation.

## 5. Pre-exponential factor

In the Reiss theory, given the total number density of clusters is small compared to the number density of single molecules of $X$ and $Y$ in the parent phase, $N_x$ and $N_y$, respectively, the pre-exponential factor $F$ of the equilibrium size distribution function in Eq. (3) is equal to the total number density of molecules

$$F = N_x + N_y. \tag{23}$$

Accordingly, in three different situations investigated by Reiss [6], it was assumed that no third inert gas was present in the parent phase. This approach was a generalization of the Frenkel model [13], which characterizes the cluster size distribution in a one-component vapour.

However, in [8], as well as in a number of earlier works (e.g. [1, 2, 14, 15]), the binary nucleation theory was applied to the generation of liquid aerosols by the homogeneous nucleation of water vapour and a polluting reactant (such as sulfuric or nitric acid) in the atmosphere. In these works, it was implicitly assumed (i.e. without justification) that Eq. (23) for the pre-exponential factor is valid also in the presence of inert carrier (atmospheric) gases.

As applied to the homogeneous nucleation of droplets from unary vapour in an inert carrier gas, such an approach was criticized by Lothe and Pound [16], who suggested that degrees of freedom corresponding to the translation of clusters have been neglected in the development of nucleation theory. As a result, they predicted that the pre-exponential factor is proportional to the number density of carrier gas molecules rather than vapour molecules, leading to a large discrepancy (many orders of magnitude) from the previous approach. A similar conclusion was made in a large number of subsequent works, reviewed and supported in [17].

This disagreement ('translation paradox') was discussed by Reiss and Katz [18], who evaluated the partition function of the system taking into account permutations of monomers among clusters and showed that Lothe and Pound's correction to the nucleation theory does not arise (for unary vapours). However, in their subsequent paper [19], where the main qualitative conclusions of [18] were reaffirmed, a correction factor of several orders of magnitude was calculated (however, much smaller than the Lothe and Pound correction). Presumably for this reason, Katz disregarded his previous results [18] and modified the Frenkel model in his subsequent works (e.g. in [20]).

Therefore, the contradiction between different approaches has not been completely resolved and required further analysis. Such an analysis for unary systems was carried out in the recent work of the author [21] within the framework of the thermodynamic approach [22], taking into account the interaction of monomers with clusters (considered in the statistical mechanics approach [18, 19] and disregarded in the Lothe and Pound model [16, 17]). The excess (or mixing) entropy calculated thermodynamically in [21] was consistent with the value evaluated in the statistical approach of Reiss, Kegel and Katz [23].

A generalization of this consideration to binary systems is presented in the Appendix, where the validity of Eq. (23) in the presence of an inert carrier gas is justified. This avoids the use of more complex expressions for binary systems from later literature (e.g. [24, 25]).

## 6. Model applications

The Reiss theory has been widely applied to the nucleation of drops in supersaturated vapour mixtures of sulfuric acid and water, e.g. in [1, 2, 14, 15]. For comparison, the modified expression for the nucleation rate, Eq. (20), can be represented in the form

$$J^* = \int_{-\infty}^{\infty} J(y')dy' = \frac{\beta_x^* \beta_y^* (1+\tan^2 \theta)}{\beta_y^* + \beta_x^* \tan^2 \theta} F \frac{|P'|}{|P''|} \left(\frac{|P''|}{Q''}\right)^{1/2} \exp\left(-\frac{\Delta G_0^*}{kT}\right), \tag{24}$$

which differs from Reiss' expression by the value of the angle $\theta$ and by the additional factor $|P'|/|P''|$.

However, as noted in [2], these parameters have only a slight effect on the nucleation rate. Moreover, the uncertainties in the literature data for other model parameters entering in the expression for $\Delta G_0^*$ are an important source of error, since the rate of nucleation has a very strong dependence upon these variables [2]. In this regard, it can be expected that the proposed modifications of the Reiss theory are not so significant and do not introduce noticeable changes comparable to the changes caused by the uncertainties of the model parameters. This conclusion was confirmed by extensive simulations of $H_2O$–$H_2SO_4$ at 22 different temperatures from 153 to 363 K with water vapour activities from 0.01 to 1.0 and sulfuric acid vapour activity from $10^{-8}$ to 1 [26]. According to 377 simulations, the difference between the mean values of $\theta$ calculated in the original and modified Reiss theory was very small, $\approx 3°$. In calculations [27] for ternary nucleation in water-ammonia-hydrocholoric acid system, the steepest descent approximation of the Rice theory was also found to be valid in a wide range of temperatures and saturation ratios.

In this situation, the Reiss theory seems to be quite satisfactory for the analysis of the available experimental data on the generation of liquid aerosols from vapour mixtures in the atmosphere, but can be modified in the course of improving the accuracy of experimental data. For more precise analysis, it is necessary to additionally take into account the effect of hydratation of the $H_2SO_4$ molecules [28], the presence of which can lead to a change in the nucleation rate [29].

## 7. Conclusion

The Reiss theory for binary homogeneous nucleation of aerosols was analysed and further developed by more accurate consideration of the steady state flow of clusters through the critical zone in the phase space of cluster sizes. For this purpose, an analysis of trajectories in the supercritical zone of the phase space was carried out within the framework of the theory of two-dimensional dynamical systems, which showed that the flow of off-critical nuclei in this zone predominates in the direction of the basic axis of the linearized system of growth rate equations, Eqs (11) and (12). From the flux matching condition at the boundary between the supercritical and critical zones, it was concluded that the flow of off-critical nuclei in the critical zone predominates in the same direction.

This substantiates the first assumption of the Reiss theory that the steady state flow of clusters through the critical zone is pronounced in one direction, but contradicts his second assumption that this direction (the axis of the pass $x'$) coincides with the direction of the steepest descent of the free energy surface. Accordingly, the modification of the Reiss theory was carried out, associated with a different direction of the axis of the pass $x'$, which should coincide with the basic axis of the linearized system.

The proposed modification is equivalent to the modification of Langer [7] and Stauffer [8], who used a more formal approach based on additional trial assumptions (ansatz) to solve the steady state equation for the non-equilibrium size distribution function in the critical zone; however, the self-consistency of these assumptions was not ultimately verified (as is usually provided for in the ansatz procedure). In our analysis, it was shown that this ansatz is not a rigorous assumption, but only an approximation (neglecting relatively small but finite deviations of the cluster trajectories from the direction of the basic axis), which can be substantiated by the flux matching condition at the boundary of the critical zone.

Although the Reiss theory was developed only for pure binary gas mixtures in the absence of a third inert gas, in subsequent works the model was applied (without additional justification) to the binary nucleation of liquid aerosols in the atmosphere using the same expression for the equilibrium size distribution function, Eqs (3) and (23), as in the Reiss theory. However, the correctness of such approach was widely criticized in the literature and therefore required additional justification, which was carried out in the present work. This avoids the use of alternative (and more complex) expressions for the pre-exponential factor in the nucleation rate from later modifications of the Reiss theory.

**Acknowledgements**

The author thanks Dr. Y. Drossinos (JRC, Ispra) for careful reading of the manuscript and valuable discussions, and Dr. V. Tarasov (IBRAE, Moscow) for useful comments and recommendations.

**Appendix. Calculation of the pre-exponential factor $F$ of the equilibrium size distribution function**

Although the Lothe and Pound approach [16] correctly identified the limitations of the earlier approach (in which the influence of an inert carrier gas was ignored), it inherited the main drawback of this approach, considering the system of monomers and clusters as an ideal mixture.

Indeed, such consideration is valid only in the case of Boltzmann statistics (to which the ideal gas obeys), when all particles are distributed over different thermodynamic states completely independently of each other [22]. For clusters of finite sizes, their interaction with monomers (described in the statistical mechanics approach [18, 19] by permutations of monomers among clusters), cannot be neglected, since clusters, in contrast to monomers, cannot be considered as point particles.

In accordance with general thermodynamics, the additivity of thermodynamic quantities, such as free energy or entropy, is preserved only as long as the interaction between different parts of the system is negligible, as in the case of ideal gas mixtures, for which, for example, the entropy of the mixture is equal to the sum of the entropies of each of gases. Therefore, for a non-ideal mixture of several substances (for example, monomers and clusters), the entropy is no longer equal to the sum of the entropies of each of the substances [22].

To find the excess entropy of a mixture of monomers and clusters at a fixed system pressure $P$, let $\Phi_0(P, T, N_a, N_b, N_0)$ be the Gibbs free energy of an ideal gas mixture (parent phase) of a carrier gas (with number density $N_0$) and vapours $A$ and $B$ (with partial pressures $P_a$, $P_b$ and number densities $N_a$, $N_b$, respectively), whose chemical potentials are $\mu_i(P, T, N_i/N) = \psi_i(P, T) + kT\ln(P_i/P) = \psi_i(P, T) + kT\ln(N_i/N)$, where $i = 0, a, b$ and $N = N_0 + N_a + N_b$. Let $\alpha_{ab}$ denote the small change which would occur in the free energy if one spherical cluster $A_{n_a}B_{n_b}$ consisting of $(n_a, n_b)$ molecules (a nucleus of the new phase) was added to the gas system. In the thermodynamic approach, clusters are considered as 'macroscopic' subsystems (or 'bodies') with $n_a, n_b \gg 1$. Due to the interactions of finite size clusters with monomers, $A_{n_a}B_{n_b} \pm A = A_{n_a \pm 1}B_{n_b}$, $A_{n_a}B_{n_b} \pm B = A_{n_a}B_{n_b \pm 1}$ (but not with inert gas molecules), this value should be sought as a function of $N_a$ and $N_b$, i.e. $\alpha_{ab} = \alpha_{ab}(P, T, N_a, N_b)$. Due to $N_{ab} \ll N_a, N_b$, where $N_{ab}$ is the number (per unit volume) of clusters of size $(n_a, n_b)$, interactions between clusters can be neglected, and thus the free energy takes the form

$$\Phi = N_0\mu_0 + N_a\mu_a + N_b\mu_b + N_{ab}\alpha_{ab}(P, T, N_a, N_b) + kT\ln(N_{ab}!), \tag{A.1}$$

where the translational entropy term, $kT\ln(N_{ab}!) \approx kTN_{ab}\ln(N_{ab}/e)$, takes into account that all (spherical) clusters of the same size $(n_a, n_b)$ are identical and, being macroscopic bodies, are homogeneously distributed in the 'external medium' (gas mixture). This is principally different from the distribution of a new ideal gas $C$ (with density $N_c \ll N \approx N_0$) in the existing gas mixture, which becomes a constituent part of the 'medium' and transforms the configurational entropy of the gases, $kT\ln\left(\frac{N!}{N_0!N_a!N_b!}\right) \approx -kT\left[N_0\ln\left(\frac{N_0}{N}\right) + N_a\ln\left(\frac{N_a}{N}\right) + N_b\ln\left(\frac{N_b}{N}\right)\right] \approx -kT\left[N_a\ln\left(\frac{N_a}{N}\right) + N_b\ln\left(\frac{N_b}{N}\right)\right]$
(which enters $\Phi$ through the chemical potential terms), into $kT\ln\left(\frac{(N+N_c)!}{N_0!N_a!N_b!N_c!}\right) \approx kT\ln\left(\frac{N!}{N_0!N_a!N_b!N_c!}\right)$, and hence the additional entropy term in Eq. (A.1) would be $kTN_c\ln\left(\frac{N_c}{N}\right)$, instead of $kTN_c \ln(N_c/e)$, with simultaneous vanishing of the interaction term $N_c\alpha_c$.

Therefore, Eq. (A.1) can be represented in the form ([22])

$$\Phi = N_0\mu_0 + N_a\mu_a + N_b\mu_b + kTN_{ab}\ln\left[\frac{N_{ab}}{e}\exp\left(\frac{\alpha_{ab}}{kT}\right)\right]. \tag{A.2}$$

Since $\Phi$ must be a homogeneous function of the first order in $N_0$, $N_a$, $N_b$ and $N_{ab}$, the term $\exp[\alpha_{ab}(P,T,N_a,N_b)/kT]$ in the argument of the logarithm (which does not depend on $N_0$) should be sought in the most general form $f_{ab}(P,T)/(N_a + \beta N_b)$. Given that after redefining $a \leftrightarrow b$, the free energy should not change, we can conclude that $\beta = 1$. Accordingly,

$$\Phi = N_0\mu_0 + N_a\mu_a + N_b\mu_b + kTN_{ab}\ln\left[\frac{N_{ab}}{e(N_a+N_b)}f_{ab}(P,T)\right], \tag{A.3}$$

or, introducing a new function $\psi_{ab}(P,T) = kT\ln f_{ab}(P,T)$,

$$\Phi = N_0\mu_0 + N_a\mu_a + N_b\mu_b + N_{ab}\psi_{ab}(P,T) + kTN_{ab}\ln\left[\frac{N_{ab}}{e(N_a+N_b)}\right]. \tag{A.4}$$

Comparison of Eq. (A.4) with Eq. (A.1) shows that

$$N_{ab}\alpha_{ab}(P,T,N_a,N_b) = N_{ab}\psi_{ab}(P,T) - kTN_{ab}\ln(N_a+N_b). \tag{A.5}$$

Therefore, since the first term in Eq. (A.5), $N_{ab}\psi_{ab}(P,T)$, does not depend on the number of monomers, the value $\psi_{ab}(P,T)$ is the standard free energy of a cluster, which for the ideal droplet model takes the form

$$\psi_{ab}(P,T) = n_a\mu_a^{(eq)} + n_b\mu_b^{(eq)} + kTn_a\ln\left(\frac{n_a}{n_a+n_b}\right) + kTn_b\ln\left(\frac{n_b}{n_a+n_b}\right) + \sigma(36\pi)^{\frac{1}{3}}(n_av_a + n_bv_b)^{\frac{2}{3}}, \tag{A.6}$$

where $v_a$ and $v_b$ are the molecular volumes of monomers and $\sigma$ is surface tension, while the second term of Eq. (A.5), $kTN_{ab}\ln(N_a+N_b)$, is the excess entropy of the mixture.

This leads to the following expressions for the chemical potentials of the 'solvents'

$$\mu'_a = \frac{\partial\Phi}{\partial N_a} = \mu_a - kTc_{ab} \approx \mu_a, \tag{A.7a}$$

$$\mu'_b = \frac{\partial\Phi}{\partial N_b} = \mu_b - kTc_{ab} \approx \mu_b, \tag{A.7b}$$

where $c_{ab} \approx N_{ab}/(N_a+N_b) \ll 1$, and of the 'solute'

$$\mu_{ab} = \frac{\partial\Phi}{\partial N_{ab}} = kT\ln c_{ab} + \psi_{ab}. \tag{A.8}$$

Therefore, from the equilibrium condition of the chemical reaction, $n_aA + n_bB = A_{n_a}B_{n_b}$,

$$n_a\mu_a + n_b\mu_b = \mu_{ab}, \tag{A.9}$$

the mass action law can be derived as

$$c_{ab} \approx N_{ab}/(N_a+N_b) = K_{ab}(T), \tag{A.10}$$

with the equilibrium constant

$$K_{ab}(T) = \exp\left(-\frac{\Psi_{ab} - n_a\mu_a - n_b\mu_b}{kT}\right) = \exp\left(-\frac{\Delta G_0(n_a, n_b)}{kT}\right), \tag{A.11}$$

where

$$\Delta G_0(n_a, n_b) = n_a\left[\mu_a^{(eq)} - \mu_a + kT\ln\left(\frac{n_a}{n_a+n_b}\right)\right] + n_b\left[\mu_b^{(eq)} - \mu_b + kT\ln\left(\frac{n_b}{n_a+n_b}\right)\right] +$$
$$\sigma(36\pi)^{1/3}(n_a v_a + n_b v_b)^{2/3} = -n_a kT\ln S_a - n_b kT\ln S_b + kT n_a \ln\left(\frac{n_a}{n_a+n_b}\right) + kT n_b \ln\left(\frac{n_b}{n_a+n_b}\right) +$$
$$\sigma(36\pi)^{1/3}(n_a v_a + n_b v_b)^{2/3} \tag{A.12}$$

is the Gibbs free energy of formation of a cluster, $S_a = N_a/N_a^{(eq)}$ and $S_b = N_b/N_b^{(eq)}$ are the supersaturations of monomers in the parent phase.

If concentrations of other clusters are also small, their contributions to the total free energy of the system are linear; therefore, the equilibrium size distribution function has the form

$$f_0(n_a, n_b) = (N_a + N_b)\exp(-\Delta G_0(n_a, n_b)/kT), \tag{A.13}$$

which is derived, as mentioned above, in the thermodynamic approach for 'macroscopic' clusters with $n_a, n_b \gg 1$. For this reason, the assertion in Ref. [23] that this expression for a cluster size of 1 does not return the number of monomers is irrelevant. Similarly, the assertion in Ref. [24] that the dependence of the concentrations of 'pure' $a$-clusters on the concentration of $b$-monomers in the gas phase $N_b$ is linear is also irrelevant, since the dependence proportional to $N_b^{n_b}$ (through $S_b$ in the exponential factor) prevails under the condition $n_b \gg 1$, even if $n_b \ll n_a$.